\documentclass[aps,prl,twocolumn,10pt,showpacs]{revtex4-1}
\usepackage{amsmath,amssymb,amsfonts}

\usepackage{inputenc}

\usepackage{tabularx}
\usepackage{graphicx}
\usepackage{amssymb,amsmath, graphicx}
\usepackage{amssymb,amsmath}
\usepackage[sort&compress]{natbib}

\begin{document}

\title{Sustainable growth in complex networks}
\author{Claudio J.~Tessone, Markus M.~Geipel, F.~Schweitzer}
\affiliation{Chair of Systems Design, D-MTEC, ETH Zurich, CH-8032 Zurich,
Switzerland}
\date{\today}

\begin{abstract}
  Based on the empirical analysis of the dependency network in 18 Java
  projects, we develop a novel model of network growth which considers
  both: an attachment mechanism and 
  the addition of new nodes with a heterogeneous distribution of their initial
  degree, $k_0$. Empirically we find that the
cumulative degree distributions of initial degrees and of the final network,
follow power-law behaviors: $P(k_{0})\propto k_{0}^{1-\alpha}$, and
  $P(k)\propto k^{1-\gamma}$, respectively. For the total number of links as a
  function of the network size, we find empirically $K(N)\propto
  N^{\beta}$, where $\beta$ is (at the beginning of the network evolution)
  between 1.25 and 2, while
  converging to $\sim 1$ for large $N$. This indicates a transition from a
  growth regime with increasing network density towards a sustainable
  regime, which prevents a collapse because of ever increasing
  dependencies. Our theoretical framework is able to predict relations
  between the exponents $\alpha$, $\beta$, $\gamma$, which also link
  issues of software engineering and developer activity. These relations
  are verified by means of computer simulations and empirical
  investigations. They indicate that the growth of real Open Source Software
networks
  occurs on the edge between two regimes, which are either dominated by
  the initial degree distribution of added nodes, or by the preferential
  attachment mechanism. Hence, the heterogeneous
  degree distribution of newly added nodes, found empirically, is
  essential to describe the laws of sustainable growth in networks.
\end{abstract}


\maketitle

How do \emph{real} networks grow? Nodes and links are not always added at
a constant rate. Instead, their numbers could be drawn from a broad
distribution, spanning almost the size of the network. This inhomogeneity
considerably impacts the network growth, but it was not covered in
existing analytical approaches. Hence, this problem is addressed in this
Letter. We provide a novel model of network growth which is solved
analytically and verified empirically by studying the evolution of
several Open Source Software projects.

% How do \emph{real} networks grow? How does a non-homogeneous mechamism
% shape a network? The processes governing real network evolution are not
% necessarily continuous. Examples where this applies can be found in many
% areas, ranging from Sociology, Engineering, Biology, Economics, etc. In
% this Letter, we unveil how does a non-continuous process modify an
% emerging network.  Also, we provide empirical evidence by studying the
% evolution of several Open Source Software projects, for which the
% complete network evolution is modified.

During its evolution, a network can have a non-linear growth of its 
set of nodes, or edges.
However, many modeling approaches, most notably the preferential attachment,
simply assume that (i) at any time step a constant number of nodes is
added to the network, that (ii) each new node is linked to the network
with a constant number of links, and that (iii) neither nodes or links
are deleted \citep{barabasi99, dorogovtsev00, krapivsky00}. If such
assumptions hold, this would result in a growth $N(\tau)\propto \tau^{\eta}$ of
the total number of nodes in the network, and $K(\tau) \propto \tau ^{\lambda}$
of the total number of links, where both $\eta\simeq 1$ and
$\lambda\simeq 1$. Such a network growth could be called
\emph{sustainable}, in contrast to the two limiting cases of (a)
accelerated growth \citep{dorogovtsev01, gagen2005, smith2009}, if
$\lambda/\eta>1$, or of (b)
saturated growth, if $\lambda/\eta<1$. Both of these growth processes are
not sustainable in the long run as they either lead to collapse or to
stagnation \citep{mattick05}.
% Indeed, empirical analyses have verified both $\eta\simeq 1$ and
% $\lambda\simeq 1$ for different kind of growing networks CITE.
But there is, at least for the intermediate observable time scales, also
empirical evidence of networks growing with increasing link density,
$K/N$ for example the World Wide Web \cite{faloutsos99}.

However, results obtained for $N(\tau)$ or $K(\tau)$ refer to macroscopic
properties, which are compatible with a large variety of `microscopic'
assumptions about node and link additions (or deletions). More
importantly, the kinetic exponents may change over time and may reach $1$
only asymptotically, which would point to changes in the growth mechanism
on intermediate time scales. Eventually, in addition to the total number
of nodes and links, there are other characteristics of the network
structure and dynamics which need to be predicted and to be verified
empirically. In this Letter, we address these problems both theoretically
and empirically by (i) developing a detailed model of network growth
which includes the heterogeneous degree distribution of newly added nodes
(instead of adding nodes with the same degree), and (ii) by verifying the
predictions of our general model against a novel data set of growing
networks.

\begin{figure*}
  \centering
\includegraphics[width=\textwidth]{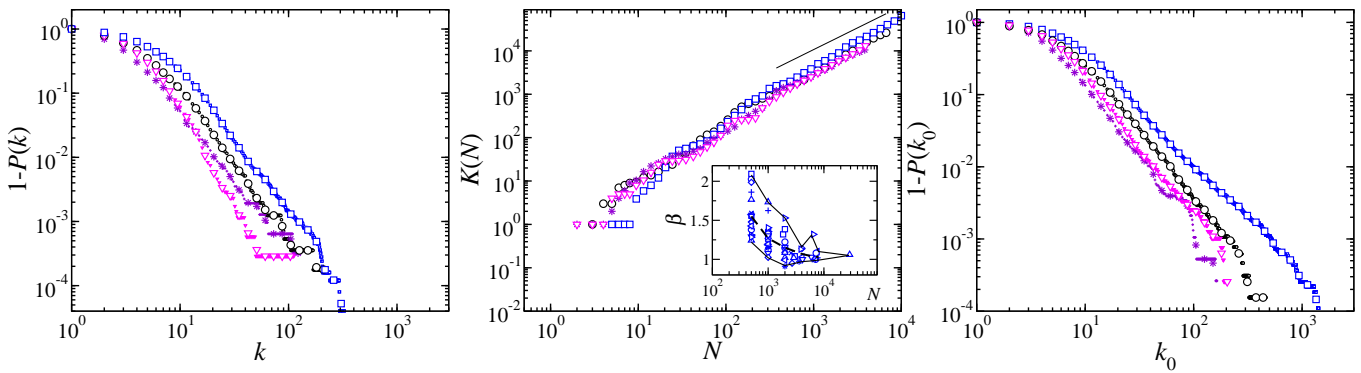}
  \caption{(Color on-line). (left) Final complementary accumulated degree 
distribution $P(k)\propto  k^{1-\gamma}$, (right) initial complementary
cumulated degree distribution
    $P(k_{0})\propto k^{1-\alpha}$, (center) total number of links
    $K(N)\propto N^{\beta}$ as a function of network size $N$. Colors
    indicate four different OSS projects: Architecturware (black circles), 
    Eclipse (blue squares), JEdit (violet stars), Sapia (magenta triangles).
See Table \ref{tab:empirical} for more details. The small symbols, represent the
complete 
    empirical datasets, while the large ones the binned data. The inset  in
the central panel shows the evolution of $\beta$ for the 18 projects, during the
growth process. In this inset, the different symbols represent the evolution of
$\beta$ for the different projects, while the dashed line represents the median
of $\beta$ for the complete set.}
  \label{fig:empirical}
\end{figure*}

We start by describing the empirical findings, to motivate the new
assumptions of our network growth model, later. We have used a dataset of
18 Open Source Software (OSS) projects (see Table 1), which are
programmed in Java. The complex network consists of \emph{nodes}, which
are Java classes (each file corresponds to one class), and \emph{directed
  links} representing dependencies between these classes. For example,
one class can call a function defined in another class, or extend a
functionality of another class. During software evolution, new classes
are added to the project and are linked to existing classes based on
principles defined in \emph{software engineering}. So, if we are able to
reveal universal dynamics underlying such growth processes, this is a
remarkable result on its own.  For the time dependent evolution of the
OSS projects, we can rely on version control systems which record all
changes made. For our analysis, we have used snapshots of intervals of 30
days, for a project life span between 2.7 and 8.2 years -- which goes
much beyond the few snapshots available for previous investigations of
OSS growth \cite{maillart08,kohring2009}. Nevertheless, we may use these
studies as a point of reference, as they also study some topological
properties, such as the cumulative degree distribution $P(k)$.

In order to derive analytical results about the latter, let us define
$n(k,\tau)$ as the degree distribution, i.e. the number of nodes with
\emph{total} degree $k$ at time $\tau$.  Obviously $K(\tau)=\sum_{k=1}^{N(\tau)}
k\, n(k,\tau)$.  The complementary cumulative degree distribution at time $\tau$
is
then
given by $P(k,\tau)=1-\sum_{l<k} n(l,\tau)/N(\tau)$. We can remove the real time
$\tau$ by using the scaling $\tau\propto N^{1/\eta}$, which means $K(N)\propto
N^{\beta}$, where $\beta=\lambda/\eta$. This procedure implies that the
number of nodes increasing, i.e. the deletion of nodes is not considered in the
empirical analysis.
Figure \ref{fig:empirical} illustrates the empirical results for these
quantities by showing four OSS projects of very different size.

Looking at the final complementary cumulative degree distribution $P(k)$,
obtained for
the maximum $N$ of the project, we clearly identify a power-law $P(k)
\propto k^{1-\gamma}$ (left panel of Fig. \ref{fig:empirical}), which is
equivalent to a degree distribution $n(k)\propto k^{-\gamma}$ The
exponents $\gamma$ which characterize the \emph{structure} of the
\emph{final product} are given in Table \ref{tab:empirical}. Dependent on
the size of the project, we find values between 2 and 3, with a clear
tendency towards values closer to 3. For the growth of the OSS projects
(center panel of Fig. \ref{fig:empirical}), we obtain slightly bend curves
for the six projects, which indicate that the exponents $\beta$ changed
over time (shown in the inset of the center panel). For every project, 
the total degree as a function of system size was split into different windows
(of size 500) and an estimation of the exponent $\beta$ was performed for each window. 
Starting at values
between 1.25 and 2, they converge to smaller values of about 1, i.e. we
observe a transition from accelerated to sustainable growth. The final
values of $\beta$ are given in Table \ref{tab:empirical}. Note that
$\beta$ is a measure of the \emph{activities of the developers}, i.e. it
characterizes a social process. The right panel of
Fig.~\ref{fig:empirical} eventually presents the most interesting
empirical finding that, different from the above mentioned assumptions
about preferential attachment and most modeling approaches, newly added nodes
have a very
heterogeneous initial degree $k_{0}$. In 	fact we observe a power-law for
the complementary cumulative \emph{initial degree} distribution
$P(k_{0})\propto
k_{0}^{1-\alpha}$, where $\alpha$ is related to the initial conditions of
the software growth, i.e. to \emph{software design}. The values found are
presented in Tab. \ref{tab:empirical}.  It remains to reveal the inherent
relations between the three exponents $\alpha$, $\beta$, $\gamma$ which
is done by the following analytical approach.

\begin{table}[htb]
  \centering
  \begin{tabular}{ p{0.25\columnwidth} >{\centering}p{0.17\columnwidth}
>{\centering}p{0.14\columnwidth} >{\centering}p{0.14\columnwidth}  c }
    \hline \hline
    Project & $N$ & $\alpha$ & $\beta$ & $\gamma$  \\
    \hline
    eclipse & 28898& $ 2.7(1) $ & $ 1.06(4) $ & $ 2.6(1) $ \\
    springframework& 7707 & $ 3.5(1) $ & $ 1.02(4) $ & $ 2.9(1) $  \\
    fudaa & 7610 & $ 2.7(1) $ & $ 1.1(1) $ & $ 2.7(1) $ \\
    jpox & 7259& $ 2.49(8) $ & $ 1.08(2) $ & $ 2.44(8) $  \\
    architecturware & 7110& $ 2.7(1) $ & $ 1.00(3) $ & $ 2.8(1) $  \\
    jena & 6619& $ 3.5(1) $ & $ 0.99(3) $ & $ 2.9(1) $  \\
    hibernate& 5938 & $ 2.5(2) $ & $ 1.03(3) $ & $ 2.5(1) $ \\
    sapia& 4129 & $ 3.44(8) $ & $ 1.00(2) $ & $ 3.0(1) $  \\
    rodin-b-sharp& 4077 & $ 2.8(1) $ & $ 1.03(2) $ & $ 2.6(1)$  \\
    azureus & 4051& $ 2.9(2) $ & $ 1.14 (5) $ & $ 2.6(2) $ \\
    jedit& 3997 & $ 2.9(1) $ & $ 1.01(1) $ & $ 2.93(8) $  \\
    jaffa& 3854 & $ 3.0(3) $ & $ 1.1(1) $ & $ 2.7(3) $  \\
    jmlspecs& 3590 & $ 2.4(2) $ & $ 0.97(6) $ & $ 2.6(2) $ \\
    openxava & 3000& $ 3.2(2) $ & $ 1.04(4) $ & $ 2.9(2) $ \\
    phpeclipse & 2881& $ 2.8(1) $ & $ 1.02(2) $ & $ 2.73(8) $ \\
    personalaccess& 2687 & $ 3.1(1) $ & $ 0.95(6) $ & $ 2.9(1) $    \\
    xmsf& 2576 & $ 2.2(1) $ & $ 1.08(3) $ & $ 2.3(1) $  \\
    aspectj & 1856& $ 2.5(1) $ & $ 1.03(4) $ & $ 2.5(1) $ \\
    \hline \hline
  \end{tabular}
  \caption{
    Empirical results obtained for 18 Open Source Java projects. $N$ gives the maximum
    number of nodes (classes) at the date of the last snapshot
    taken; $\alpha$,  $\gamma$ are the exponents  for the
    initial and final degree distribution. $\beta$ is the 
    value of the exponent describing the asymptotic growth of the total number
of links as a function of network size.  
  }
  \label{tab:empirical}
\end{table}

\begin{figure}[t]
  \centering
  \includegraphics[width=0.4\textwidth]{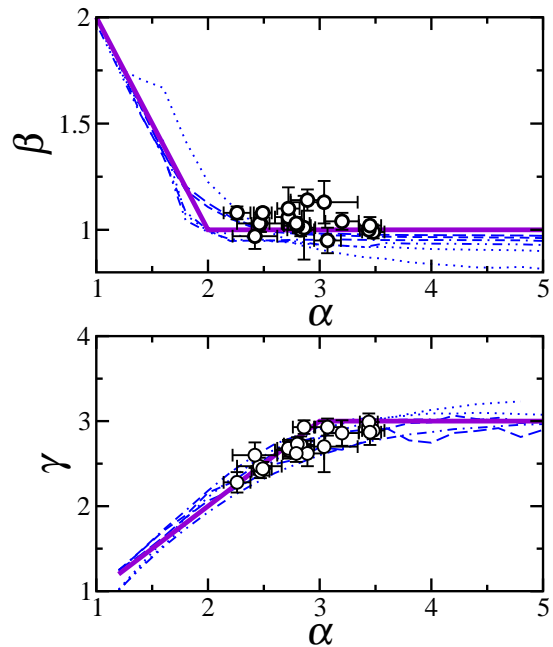}
  \caption{(upper panel) Exponents $\gamma$ of the power-law degree
    distribution of the final network,  (lower panel) exponents $\beta$ for the
growth
    of the total number of links as a function of the exponent
    $\alpha$ of    the initial degree distribution.  
    The different thin lines correspond to simulations of
    the model described, for various network sizes ($N=2\times 10^3$ --dotted
line-- to
    $N= 10^5$ --dashed line--). The thick lines indicate the analytical results
of
    Eqs.~(\ref{eq:beta}) and (\ref{gamma}). Marks with error bars correspond
    to the empirical results for the 18 projects, given in Table
\ref{tab:empirical}.   }
  \label{fig:dd_exponents}
\end{figure}

We assume that nodes are added to the project at a constant rate,
i.e. time $t$ is given by the total number of nodes, $t=N$, or conversely,
$\eta=1$. For the dynamics of the degree distribution we
postulate the following rate equation:
\begin{eqnarray}
  \dot n(k,t) &=& \delta_{k,k_{0}(t)} 
  + n(k-1,t) \, \omega[ k-1 \to k ] \nonumber \\ & & 
  + n(k+1,t) \, \omega[ k+1 \to k ]  \\
  & & - n(k,t)\, \big\{ \omega[ k \to k-1 ]  
  + \omega[ k \to k+1  ] \big\} \nonumber
  \label{eq:rates} 
\end{eqnarray}
This is a first order approximation of the dynamics based on the
addition/deletion of one node at a time.  The term $\delta_{k,k_{0}(t)}$
in Eq.~(\ref{eq:rates}) describes the addition of a new node with an
initial degree exactly equal to $k$. In accordance with our empirical
findings, this degree is randomly drawn from a truncated power-law
distribution $g(k_{0})$ with exponent $\alpha$; i.e.~$ \hbox{Prob}[k_0(t)
= k] = \min \left( (\alpha-1)/k^\alpha, t-1 \right)$. For the transition
rate of growth processes, $k \to k+1$, we assume
\begin{equation}
  \omega[ k \to k+1] = \left\{\frac{ k_0(t) }{K(t)} 
+ \left(\sigma+\frac{r}{2}\right) \right\} k. \label{eq:rates:plus}
\end{equation}
This rate is proportional to $k$, i.e. it is based on preferential
attachment. Without that assumption, the process would result in a
single-scale network which is not in accordance with the empirical
studies above. The preferential attachment can occur by means of two
different processes: The first one occurs if a newly added node with 
$k_0(t)$ new links to existing nodes, which are selected with a
probability proportional to their relative degree $k/K$. The second
process describes the addition of links between existing nodes, where
$\sigma$ and $r$ are constants described below.  The transition rate
corresponding to the deletion of links, $k \to k-1$, is also assumed to
be proportional to the degree of the node,
\begin{equation}
  \omega[ k \to k-1] = 
  \left\{ \sigma - \frac{r}{2} \right\} k.\label{eq:rates:minus}
\end{equation}

Let us now elucidate the emerging dynamics by splitting it into two different
limiting cases. The first one 
occurs when the growth of the network based on the \emph{addition of
  nodes} with heterogeneous degree $k_{0}$, does not play any role.
I.e. $k_{0}$ can be set to zero, for every time step. In this case the dynamics
is only
governed by the \emph{addition/deletion of links} distributed between
existing nodes, which follows the preferential attachment/deletion rule.
Then, the rate equation (\ref{eq:rates}), in the continuous limit and for
large $N$, can be easily translated into the following Fokker-Planck
equation:
\begin{equation}
\partial_t n(k,t) = r \, k \, n(k,t) + \partial^2_{kk} 
\left(  \sigma^2\, k^2 \right) \, n(k,t) 
\label{eq:fpe}
\end{equation}
which is equivalent to the following Langevin dynamics for the degree
$k_{i}$ of a single node $i$:
\begin{equation}
\dot k_i(t) = - r\, k_i(t) + \sigma \,  k_i(t) \,\xi_i(t),\label{eq:lang}
\end{equation}
This describes the known \emph{law of proportional growth}
\cite{simon58,saichev09}, where $r$ is the mean growth (drift) and $\sigma$ is
the
variance of the normalized random force $\xi_i(t)$. It is well known
\cite{clauset09} that such processes in the long run lead to a power-law
distribution $n(k)\propto k^{-\gamma}$, i.e. to Zipf's law for the
cumulative degree distribution $P(k)\propto k^{1-\gamma}$, with $\gamma$
equal to 2. In fact, Zipf's law was empirically confirmed for the
\emph{in-degree} distribution of \emph{Linux packages} \cite{maillart08}
as well as for \emph{Java projects} \cite{kohring2009}. However, the
\emph{out-degree} distribution, at least for the latter dataset, clearly
follows a \emph{log-normal} distribution. After all, because this limit
case only considers the growth of the number of links, but not of the
number of nodes, it only provides a limited understanding of real
software dynamics (and random addition/deletion processes are only one of
many different ways to obtain Zipf's law).

Therefore, in this Letter, we are more interested in the second limiting
case which ignores the addition/deletion of links among existing nodes --i.e.
$\sigma$, $r$ are negligible--,
while emphasizing the network growth based on the
addition of nodes with a broad initial degree distribution,
$g(k_{0})$. This assumption is fully justified in the case of broad
distributions of initial degrees, as found empirically. This dynamics is fully
described by the following set of
equations:
\begin{eqnarray}
  \dot n(1,t)&=& \delta_{1,k_0(t)} - \frac{k_0(t) }{K(t)} n(1,t) \label{eq:evol1} \\
  \dot n(k,t)&=&  \delta_{k,k_0(t)} 
+ \frac{k_0(t)}{K(t)} \big\{ (k-1) n(k-1,t)- k\, n(k,t) \big\} 
\nonumber 
\end{eqnarray}
and the initial condition $n(k,0)= n_0 \, \delta_{k,n_{0}-1}$.
I.e. initially a small number of nodes (e.g. $n_{0}=2$) with a degree of
$n_{0}-1$ is assumed, which describes a small, fully connected network to
start with. From this set of equations, we first derive the dynamics for
the total number of links, $K(t)$. 
By definition, for a single network realization $ \dot K(t) = k_0(t)$
holds.  The ensemble average $\langle K(t) \rangle$ over many
realizations of the network growth process is then given by:
\begin{equation}
  \left\langle \dot K(t) \right\rangle = 
\left\langle k_0 | k_0 < t \right\rangle
  + t \cdot
  \hbox{Prob}[k_0(t) > t].
\label{eq:kdot0}
\end{equation}
The first term 
represents the expected value of $k_0(t)$ restricted to $k_0(t) < t$,
which applies if the number drawn from the distribution $g(k_0)$ is lower
than the current network size ($t=N$) and, thus, the newly added node is
able to establish as many links as drawn from the distribution. If this
is not the case, i.e.~$k_0(t)>t$ the node can only create at most $t-1$
links, which is described by the second term.  By recasting the power-law
distribution for $g(k_0)$, we get after some calculation:
\begin{equation}
  \left\langle \dot K(t) \right\rangle =\int\limits_1^t dk\, g(k) \,k + t\, \int\limits_t^\infty
  dk \,g(k)\ 
= \frac{\alpha-1}{2-\alpha} + \frac{t^{2-\alpha}}{2-\alpha}. \label{eq:kdot}
\end{equation}
Asymptotically, we find that the total number of links grows in time or
with network size $t=N$, respectively,  as a power-law, $K(t) \propto
t^\beta$, with the exponent
\begin{equation}
\beta  = 
\begin{cases}
  3-\alpha & \hbox{if}  \,\, \,\, \alpha < 2 \\
  1 & \hbox{if} \,\, \,\,   \alpha \geq 2 \\
\end{cases}. \label{eq:beta}
\end{equation}

By applying the ensemble average to Eqs.~(\ref{eq:evol1}),
we are further able to find a mean-field approximation for the dynamics
of the degree distribution $n(k,t)$. 
Using $\langle \delta_{k,k_0(t)} \rangle = \hbox{Prob}[k = k_0(t)] =
(\alpha - 1) / k^\alpha$ and similar arguments as in
Eqs.~(\ref{eq:kdot}-\ref{eq:beta}), we find that
\begin{equation}
\langle k_0(t) \rangle =  t^{2-\alpha} \frac 1 {2-\alpha}  +
\frac{\alpha-1}{\alpha-2}. \label{eq:k0}
\end{equation}

By analyzing the solution of Eqs.~(\ref{eq:kdot})-(\ref{eq:k0}) we find
two different regimes for the ratio $\langle k_0(t) \rangle / \langle
K(t) \rangle $: (i) if $\alpha>2$, then $\langle k_0(t) \rangle \propto
(\alpha-1)/(\alpha-2)$ and $\langle K(t) \rangle \propto
(\alpha-1)/(\alpha-2)$, (ii) if $\alpha<2$, $\langle k_0(t) \rangle
\propto t^{2-\alpha}/(\alpha-2)$ and $\langle K(t) \rangle \propto
t^{3-\alpha}$. Both regimes, however, yield identical result,
i.e.~$\langle k_0(t) \rangle / \langle K(t) \rangle = \zeta(\alpha) t$,
with $\zeta (\alpha)$ being a normalization constant. Thus, we can
rewrite Eqs.~(\ref{eq:evol1}) as,
\begin{eqnarray}
  \langle \dot n(1,t) \rangle  &=& (\alpha - 1) -  \frac{\langle  n(1,t)
\rangle}{ \zeta(\alpha) \, t} \\
  \langle \dot n(k,t) \rangle  &=& \frac{(\alpha - 1)}{ k^\alpha }+  \frac{(k-1)
\langle n(k-1,t) \rangle- k \langle n(k,t) \rangle}{ \zeta(\alpha) \, t}
\nonumber
\end{eqnarray}
These equations reveal a competition between two different processes: the
growth of the network caused by the addition of links with a broad initial
degree distribution  (first term) and the growth of a node's degree caused by
a mechanism akin to preferential attachment (second term). If $\alpha$ is
small, the first
case dominates and
the expected degree distribution is simply given by $\langle \dot n(k,t)
\rangle = t{(\alpha - 1)}/{ k^\alpha }$. 
On the other hand, if $\alpha$ is large and the addition of new nodes
with a heterogeneous initial degree distribution can be neglected, we
recover the usual Barab\'asi-Albert model with $n(k,t)\propto
k^{-3}$. 
Thus, we have found two different regimes for the final
degree distribution, which depend of the exponent of the initial degrees
distribution $\alpha$:
\begin{equation}
\gamma  = 
\begin{cases}
  \alpha & \hbox{if}  \,\, \,\, \alpha < 3 \\
  3 & \hbox{if} \,\, \,\,   \alpha \geq 3 \\
\end{cases}.
\label{gamma}
\end{equation}
To conclude, our analytical approach has provided a firm relation between
the three different exponents $\alpha$, $\beta$, $\gamma$, which can be
tested in two different ways: (a) by computer simulations of the full
dynamics for various network sizes $N$ and initial conditions
($\alpha$), 
(b) by comparison with the empirical findings from the 18 OSS
projects. The results are shown in Figure \ref{fig:dd_exponents}. They
confirm that the analytical approximations are indeed valid and in good
agreement both with the computer simulations and the empirical
results. Most interestingly, they reveal that the growth dynamics of real
OSS networks is on the edge between two regimes: for $\alpha<3$, the
initial degree distribution and hence the addition of new nodes would
dominate the whole growth process, whereas for $\alpha>3$ the
preferential attachment of links between existing nodes would
dominate. As the empirical findings verify, none of these regimes fully
cover real software growth. In particular, the heterogeneous degree
distribution of newly added nodes cannot be neglected.

Eventually, we wish to point to the self-organizing dynamics observed in
OSS, which turns an initially accelerated network growth ($\beta>1$) into
a sustainable one ($\beta \to 1$) found for mature projects. This
prevents a collapse of the software growth due to non-linearly increasing
dependencies between classes. Interestingly, $\beta$, which describes the
effort (social activity) of developers adding new classes to the
software, is found to be closely related to the other two exponents
$\alpha$, $\gamma$, describing a very different `dimension' of the
software evolution, namely software engineering. This may shed new light
on the underlying principles of software design and software project
management.

\bibliographystyle{apsrev4-1}
\bibliography{refs}

\end{document}